\documentclass[reprint,final]{revtex4-1}

\usepackage{amssymb} 
\usepackage{amsthm}
\usepackage{amsmath}
\usepackage{graphicx,subfigure}
\usepackage{dcolumn}
\usepackage{bm}
\usepackage{cancel}
\usepackage{float}
\usepackage{epstopdf}
\usepackage{hyperref}
\usepackage{url}
\usepackage[svgnames] {xcolor}

\hypersetup{colorlinks=true, linkcolor=blue, urlcolor=blue, citecolor=blue}

\begin{document}

\title{Reactor Antineutrino Signals at Morton and Boulby}

\author{S.T. Dye}
\affiliation{Department of Physics and Astronomy, University of Hawaii, Honolulu, HI, 96822 USA}
\affiliation{Department of Natural Sciences, Hawaii Pacific University, Kaneohe, HI, 96744 USA}

\date{\today}

\begin{abstract}
\vspace{1mm}
\noindent
Increasing the distance from which an antineutrino detector is capable of monitoring the operation of a registered reactor, or discovering a clandestine reactor, strengthens the Non-Proliferation of Nuclear Weapons Treaty. This report presents calculations of the reactor antineutrino interactions, from both quasi-elastic neutrino-proton scattering and elastic neutrino-electron scattering, in a water-based detector operated $\gtrsim10$ km from a commercial power reactor. It separately calculates signal from the proximal reactor and background from all other registered reactors. The main results are the interaction rates and kinetic energy distributions of the charged leptons scattered from the quasi-elastic and elastic processes. Comparing signal and background distributions evaluates reactor monitoring capability. Scaling the results to detectors of different sizes, target media, and standoff distances is straightforward. Calculations are for two specific examples of a commercial reactor ($P_{th}\sim3$ GW) operating nearby ($L\sim20$ km) an underground facility capable of hosting a detector ($\sim1$ kT H$_2$O) project. These reactor-site combinations are Perry-Morton on the southern shore of Lake Erie in the United States and Hartlepool-Boulby on the western shore of the North Sea in England. The signal from the proximal reactor is about five times greater at the Morton site than at the Boulby site due to shorter reactor-site separation distance, larger reactor thermal power, and greater neutrino oscillation survival probability. In terms of absolute interaction rate, background from all other reactors is larger at Morton than at Boulby. However, the fraction of the total rate is smaller at Morton than at Boulby. Moreover, the Hartlepool power plant has two cores whereas the Perry plant has a single core. These conditions make monitoring the operation cycle of a nuclear reactor more challenging at the Boulby site than at the Morton site. The Boulby site, therefore, offers an opportunity for demonstrating remote reactor monitoring under more stringent conditions than does the Morton site.
\end{abstract}

\maketitle

\section{introduction} 
Monitoring the operation of a reactor from a remote location through the detection of antineutrinos is a nuclear nonproliferation goal \cite{adamb10}. Nuclear monitoring activities and studies primarily utilize quasi-elastic antineutrino-proton scattering, commonly called the inverse beta decay reaction \cite{bowden09,snif10,nudar13}. Coincidence counting of both reaction products of 
\begin{equation}
\overline{\nu}_e+p \rightarrow e^+ + n,
\label{ibd}
\end{equation}
a positron and a neutron, is a traditional technique \cite{reines53}, which drastically reduces background. While the angular distribution of the positrons is nearly isotropic in the energy range of reactor antineutrinos, the neutrons scatter in the forward direction \cite{vogel99}. Resolving the direction to the source of antineutrinos using inverse beta decay relies on measuring the direction of the outgoing neutron. Although the uncharged neutron does not directly produce a trail of ionization, the location of its capture is generally farther from the antineutrino source than the positron track. Asymmetry in the ensemble of positions of neutron capture relative to positron production is apparent in the data of several reactor antineutrino detection experiments, using scintillating liquid \cite{chooz00,paloverde00}. An additional reaction to employ for the remote monitoring of nuclear reactors is elastic neutrino-electron scattering. This reaction 
\begin{equation}
\overline{\nu}_l+e^- \rightarrow \overline{\nu}_l+e^-
\label{es}
\end{equation}
$(l=e,\mu,\tau)$, whether induced by a neutrino or an antineutrino, always knocks the electron into the hemisphere directed away from the source. The scattered electron ionizes the detection medium and, if sufficiently energetic, produces Cherenkov radiation in a cone around the direction of the electron track. Reconstructing the ionization trail or the Cherenkov ring estimates track direction, which rejects background. Monitoring nuclear fusion in the Sun using directional Cherenkov radiation from elastic neutrino-electron scattering in water is well established \cite{kamioka90,sk98,sno01}. Adding information on reactor antineutrino direction to flux and spectral information enhances nuclear monitoring capabilities \cite{nudar13}. Water doped with gadolinium, which facilitates the detection of neutron captures, enhances the opportunity to combine flux and spectral information from quasi-elastic neutrino-proton scattering with directional information from elastic antineutrino-electron scattering.

This report presents calculations of reactor antineutrino interactions in a distant detector. These calculations are the foundation for assessing the capability of remotely monitoring the operation of a nuclear reactor. Scaling the results to detectors of different sizes, target media, and standoff distances is straightforward. Reliable measurement of antineutrino signals from a remote reactor depends critically on background and detector sensitivity \cite{hellfeld}. This report estimates background from other reactors but does not yet include background from geo-neutrinos, solar neutrinos, cosmogenic radionuclides, or detector noise, including radon contamination, nor does it presently consider detector sensitivity. These considerations are topics of future additions to this report. The next addition is probably background from geo-neutrinos.

\section{Reactor-Site Combinations}
There are two unique combinations of a commercial reactor ($P_{th}\sim3$ GW) operating nearby ($L\sim20$ km) an underground facility capable of hosting a detector ($\sim1$ kT H$_2$O) project. These reactor-site combinations are Perry-Morton on the southern shore of Lake Erie in the United States and Hartlepool-Boulby on the western shore of the North Sea in England. Both the Perry-Morton and the Hartlepool-Boulby combinations are under consideration for demonstrating remote monitoring of a nuclear reactor through the detection of antineutrinos \cite{watchman}. Table~\ref{tab:sites} lists the thermal power $P_{th}$, type and number of cores of the proximal nuclear reactor plant \cite{infnsite}, reactor-site separation distance $L$, and overburden $D$ for these combinations. Overburden is important for estimating cosmogenic background \cite{hellfeld}. Background sources are other reactors, solar neutrinos, geo-neutrinos, cosmogenic nuclides, neutrons, and detector noise, including radon contamination. This report presently considers background only from other reactors. 
\begin{table}
\caption{The thermal power $P_{th}$, type and number of cores, standoff distance $L$, and overburden $D$ for the Perry-Morton and Hartlepool-Boulby reactor-site combinations.}
\begin{tabular}{l c c c c c}
\hline\noalign{\smallskip}
  & $P_{th}$(MW) & Type & Cores &$L$(m) & $D$(m.w.e.) \\
\hline\noalign{\smallskip}
Perry-Morton & $3758$ & BWR & 1 & $13000$ & $1560$ \\
 \noalign{\smallskip}
Hartlepool-Boulby & $3000$ & GCR & 2 & $25000$ & $2800$ \\
\hline\noalign{\smallskip}
\end{tabular}
\label{tab:sites}
\end{table}

\section{Reactor Antineutrino Signal Spectrum}
\begin{figure}
\centering
\includegraphics[trim = 5mm 40mm 15mm 40mm, clip, scale=0.45]{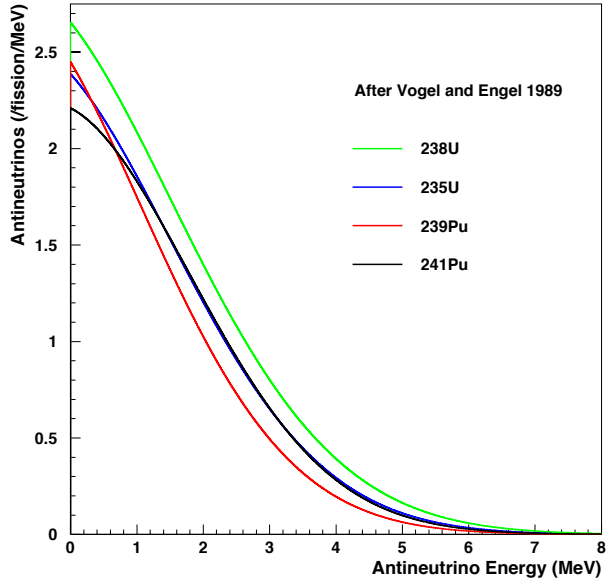}
\caption{Antineutrino energy spectra for the four main fission isotopes, $^{235}$U, $^{238}$U, $^{239}$Pu, and $^{241}$Pu, in a nuclear reactor.}
\label{fig:spectra}
\end{figure}

Nuclear reactors generate heat by fissioning uranium and plutonium isotopes. The four main isotopes are $^{235}$U, $^{238}$U, $^{239}$Pu, and $^{241}$Pu. Antineutrinos emerge from the beta decay of the many fission fragments. An estimate of the energy spectrum of antineutrinos sums the exponential of a degree two polynomial of antineutrino energy ($E_\nu$) for each of the main isotopes \cite{vogel_engel}. Specifically,
\begin{equation}
\lambda(E_\nu)=\mathrm{exp}(a_0+a_1E_\nu+a_2E_\nu^2),
\end{equation}
where the coefficients $a_{j=0,1,2}$ are fit parameters. Figure~\ref{fig:spectra} shows the estimated spectra for the four main isotopes. Each isotope ($i=1, 2,3,4$) contributes a fraction of the reactor power ($p_i$), releasing an average energy per fission ($Q_i$). The estimated energy spectrum of the reactor antineutrino emission rate is \cite{baldoncini}
\begin{equation}
dR/dE_\nu=P_{th}\sum_i \frac{p_i}{Q_i} \lambda_i(E_\nu).
\end{equation}
Table~\ref{tab:rspec} lists the power fraction \cite{bellini10} and energy per fission \cite{ma13} values used to estimate the reactor spectrum. This report assumes that both boiling water reactors (BWR) and gas cooled reactors (GCR) produce the same energy spectrum of antineutrinos. 
\begin{table}
\caption{Reactor spectrum power fractions $p_i$ and fission energies $Q_i$ for the four main isotopes.}
\begin{tabular}{l c c c c}
\hline\noalign{\smallskip}
                     & $^{235}$U & $^{238}$U & $^{239}$Pu & $^{241}$Pu \\
\hline\noalign{\smallskip}
$p_i$            & .56 & .08 & .30 & .06 \\
$Q_i$ (MeV) & 202.4 & 206.0 & 211.1 & 214.3 \\
\hline\noalign{\smallskip}
\end{tabular}
\label{tab:rspec}
\end{table}

\section{Neutrino Oscillations}
Neutrino flavors ($e$, $\mu$, $\tau$) are quantum mechanical mixtures of three neutrino mass states ($m_1$, $m_2$, $m_3$). Mixture varies with distance travelled as a function of energy, according to the well established phenomenon of neutrino oscillations. The probability that an electron antineutrino of energy $E_\nu$ in MeV loses its flavor after traveling a distance $L$ in meters is
\begin{equation}
\label{nuosc}
\begin{split}
P_{e\rightarrow\mu,\tau}(L,E_\nu)=\cos^4\theta_{13}\sin^22\theta_{12}\sin^2\Delta_{21}\\
+\cos^2\theta_{12}\sin^22\theta_{13}\sin^2\Delta_{31}\\
+\sin^2\theta_{12}\sin^22\theta_{13}\sin^2\Delta_{32},
\end{split}
\end{equation}
where $\Delta_{ij}=1.27(|\delta m_{ji}^2|L)/E_\nu$ with $\delta m_{ji}^2=m_j^2-m_i^2$ the neutrino mass-squared difference in eV$^2$ and $\theta_{12}$, $\theta_{13}$ are the solar, reactor mixing angles, respectively. The complimentary probability, $P_{e\rightarrow e} = 1 - P_{e\rightarrow\mu,\tau}$, gauges survival of electron flavor. Table~\ref{tab:nuosc} lists the neutrino oscillation parameter values \cite{capozzi16} used to estimate the spectra of detectable reactor antineutrino interactions, assuming normal mass ordering ($m_3>m_2>m_1$).
\begin{table}
\caption{Neutrino oscillation parameter values as functions of the solar mixing angle $\theta_{12}$, reactor mixing angle $\theta_{13}$, and mass-squared differences.}
\begin{tabular}{c c c c}
\hline\noalign{\smallskip}
$\sin^2 \theta_{12}$  & $\delta m_{21}^2$ & $\sin^2 \theta_{13}$ & $\delta m_{31}^2$ \\
\hline\noalign{\smallskip}
.297 & $7.37\times10^{-5} \mathrm{eV}^2$  & .0214 & $2.50\times10^{-3} \mathrm{eV}^2$ \\
\hline\noalign{\smallskip}
\end{tabular}
\label{tab:nuosc}
\end{table}

Reactor antineutrinos which oscillate to $\overline\nu_{\mu}$ and $\overline\nu_{\tau}$ with probability complimentary to \eqref{nuosc} do not initiate inverse beta decay \eqref{ibd}. They do, however, interact by elastic scattering \eqref{es} although with smaller cross section than do $\overline\nu_e$. Neutrino oscillations reduce the interaction rate and distort the energy spectrum of the detected charged lepton more for \eqref{ibd} than for \eqref{es}. The spectral distortion of \eqref{ibd} interactions provides important information on the distance to the source of antineutrinos \cite{dye09,nudar13}.

\section{Antineutrino Scattering}
The scattering of reactor antineutrinos in ordinary matter primarily occurs by two processes. The dominant reaction is quasi-elastic neutrino-proton scattering \eqref{ibd}, or inverse beta decay (IBD). The reaction cross section follows from the V-A theory of weak interactions \cite{vogel99}. Neglecting energy-dependent recoil, weak magnetism, and radiative corrections, the cross section is
\begin{equation} 
\label{pxsec}
\sigma^{IBD} (E_e) = \sigma^{IBD}_0 p_e E_e,
\end{equation}
where $E_e$ and $p_e= \sqrt{E^2_e-m^2_e}$ are the positron energy and momentum, respectively, and $m_e$ is the positron mass. Assuming the nucleon mass is infinite, then the energy of the incident neutrino $E_\nu$ relates to the energy of the positron $E_e$ by
\begin{equation}
E_\nu = E_e + \Delta,
\end{equation}
where $\Delta = M_n - M_p$ is the neutron proton mass difference \cite{vogel99}.

The energy independent coefficient $\sigma^{IBD}_0$, which contributes the dominant error in the evaluation of $\sigma^{IBD}$, depends on experimental data. Input comes from either the free neutron lifetime $\tau_n$ or the axial-to-vector coupling ratio $\lambda  = |g_A/g_V|$. Normalizing the cross section to the $\beta$-decay of the free neutron gives 
\begin{equation}
\sigma^{IBD}_0 = \frac{2\pi^2} {m_e^5f^R\tau_n},
\label{nlife}
\end{equation}
where $f^R$ is the phase space factor \cite{vogel99}. Normalizing the cross section to the axial-to-vector coupling ratio gives
\begin{equation}
\sigma^{IBD}_0 = \frac{G_F^2 cos^2\theta_C}{\pi} (1+\Delta_{inner}^R)(1+3\lambda^2),
\label{coupling}
\end{equation}
where $G_F$ is the Fermi constant, $\theta_C$ is the Cabibbo angle, and $\Delta_{inner}^R $ accounts for the inner radiative corrections \cite{vogel99}. Using standard values both \eqref{nlife} and \eqref{coupling} give a value consistent with $\sigma^{IBD}_0 = 9.62 \times10^{-44}$ cm$^2$/MeV$^2$. 

Standard electroweak theory gives the cross section for the sub-dominant process of neutrino-lepton scattering. For electron antineutrinos scattering on electrons \eqref{es} the total cross section is \cite{fukuyana}
\begin{equation}
\label{exsec}
\begin{split}
\sigma^{ES}_{\overline{\nu}_e}(E_\nu)=\frac{G_F^2m_e}{6\pi}E_\nu[(1+4\sin^2\theta_W+16\sin^4\theta_W)\\
-(3\sin^2\theta_W+6\sin^4\theta_W)\frac{m_e}{E_\nu}],
\end{split}
\end{equation}
where $G_F$ is the Fermi coupling constant, $m_e$ is the electron mass, and $\theta_W$ is the weak mixing angle. Evaluating the energy independent coefficient using standard values gives
\begin{equation}
\sigma^{ES}_0 = \frac{G_F^2m_e}{6\pi} = 1.436\times10^{-45} \mathrm{cm}^2/\mathrm{MeV}.
\end{equation}

At high energy ($E_\nu \gg m_e$) terms involving $m_e/E_\nu$ in \eqref{exsec} are negligible. This leads to an approximate form of the total cross section
\begin{equation}
\label{apprx}
\sigma^{ES}_{\overline{\nu}_e}(E_\nu)\cong \sigma^{ES}_0E_\nu(1+4\sin^2\theta_W+16\sin^4\theta_W).
\end{equation}
Use of the approximate cross section \eqref{apprx} overestimates the number of reactor antineutrino elastic scattering interactions, especially at low energy.

The detectable particle in \eqref{es} is the electron, which always scatters forward relative to the direction of the neutrino. Conservation of energy and momentum define the scattering angle $\theta$  in terms of the electron kinetic energy $T_e$,
\begin{equation}
\cos\, \theta=\frac{1+m_e/E_\nu}{(1+2m_e/T_e)^{1/2}}.
\end{equation}
The electron kinetic energy is maximum for scattering in the direction of the incident neutrino (cos$\, \theta=1$),
\begin{equation}
T_{e,\mathrm{max}}=\frac{E_\nu}{(1+m_e/2E_\nu)}.
\end{equation}
When evaluating the detectable signal it is useful to consider the differential cross section
\begin{equation}
\frac{d\sigma_{\overline{\nu}_e}}{dT_e}=3\sigma^{ES}_0[A_0+B_0(1-\frac{T_e}{E_\nu})^2+C_0\frac{m_eT_e}{E_\nu^2}],
\end{equation}
with the coefficients $A_0$, $B_0$, and $C_0$ given in terms of the weak mixing angle $\theta_W$. Specifically, $A_0=4\sin^4\theta_W$, $B_0=(A_0+1+4\sin^2\theta_W)$, and $C_0=-(A_0+2\sin^2\theta_W)$ \cite{fukuyana}. Setting $C_0=0$ gives the approximate form of the differential cross section, corresponding to \eqref{apprx}.

Reactor antineutrinos that oscillate from $\overline{\nu}_e$ to $\overline{\nu}_{\mu,\tau}$ lack the energy to create a $\mu^+$ or $\tau^+$ and therefore do not induce quasi-elastic scattering. The resulting distortion of the energy spectrum detected by \eqref{ibd} reveals the full effect of neutrino oscillations. The same $\overline{\nu}_{\mu,\tau}$ do induce elastic scattering \eqref{es}, although with smaller interaction cross section than $\overline{\nu}_e$, thereby recovering some of the signal. The total cross section for $\overline{\nu}_{\mu,\tau}+e^- \rightarrow \overline{\nu}_{\mu,\tau}+e^-$ is
\begin{equation}
\label{muxsec}
\begin{split}
\sigma^{ES}_{\overline{\nu}_{\mu,\tau}}(E_\nu)=\sigma^{ES}_0E_\nu[(1-4\sin^2\theta_W+16\sin^4\theta_W)\\
-(3\sin^2\theta_W+6\sin^4\theta_W)\frac{m_e}{E_\nu}].
\end{split}
\end{equation}
The differential cross section is
\begin{equation}
\begin{split}
\frac{d\sigma_{\overline{\nu}_{\mu,\tau}}}{dT_e}=3\sigma^{ES}_0[A_0+(B_0-8\sin^2\theta_W)(1-\frac{T_e}{E_\nu})^2\\
+(C_0+4\sin^2\theta_W)\frac{m_eT_e}{E_\nu^2}].
\end{split}
\end{equation}

\begin{figure}
\centering
\includegraphics[trim = 5mm 40mm 15mm 40mm, clip, scale=0.45]{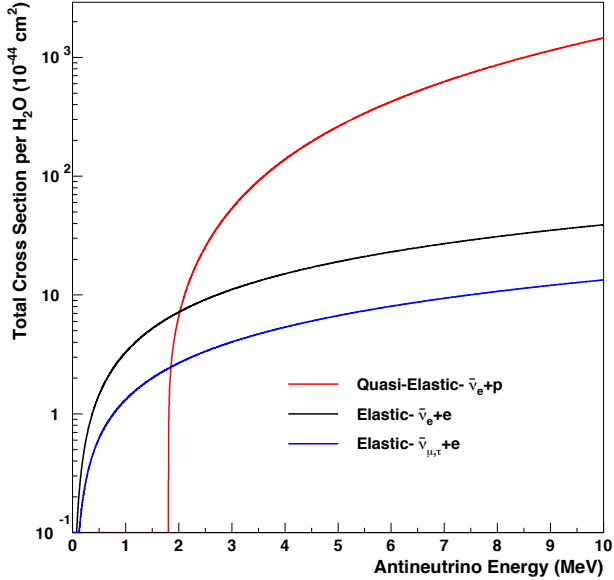}
\caption{Quasi-elastic and elastic scattering total cross sections as a function of antineutrino energy.}
\label{fig:sigma}
\end{figure}

Figure~\ref{fig:sigma} shows the total cross section per water molecule for \eqref{pxsec}, \eqref{exsec}, and \eqref{muxsec} over the energy range relevant to the reactor spectrum ($0-10$ MeV). Whereas \eqref{pxsec} is largest for $E_\nu>2$ MeV, unlike \eqref{exsec} and \eqref{muxsec} it provides no sensitivity at energy below $1.8$ MeV.

\section{Interaction Rates}
\begin{figure}
\centering
\includegraphics[trim = 5mm 40mm 15mm 40mm, clip, scale=0.45]{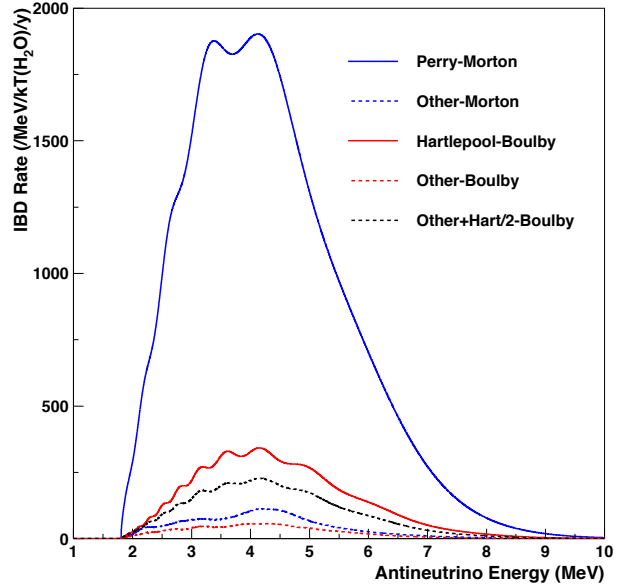}
\caption{Energy spectra for quasi-elastic scattering (IBD) interactions from the proximal reactor at the Morton and Boulby sites. The dashed black curve sums the integral rate of quasi-elastic interactions from one of the cores at Hartlepool (Hart/2) and all other reactors at the Boulby site.}
\label{fig:ibd_rate}
\end{figure}

\begin{figure}
\centering
\includegraphics[trim = 5mm 40mm 15mm 40mm, clip, scale=0.45]{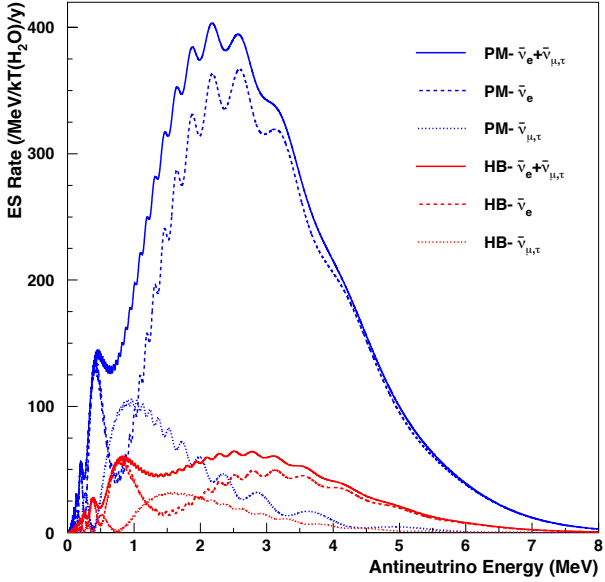}
\caption{Energy spectra for elastic scattering (ES) interactions from the proximal reactor at the Morton (PM) and Boulby (HB) sites.}
\label{fig:es_rate}
\end{figure}

\begin{figure}
\centering
\includegraphics[trim = 5mm 40mm 15mm 40mm, clip, scale=0.45]{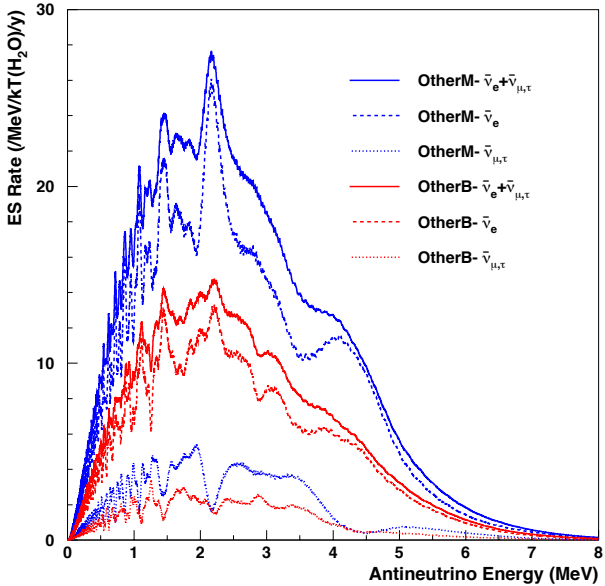}
\caption{Energy spectra of elastic scattering (ES) interactions from all other reactors at the Morton (OtherM) and Boulby (OtherB) sites.}
\label{fig:es_rate_other}
\end{figure}

\begin{figure}
\centering
\includegraphics[trim = 5mm 40mm 15mm 40mm, clip, scale=0.45]{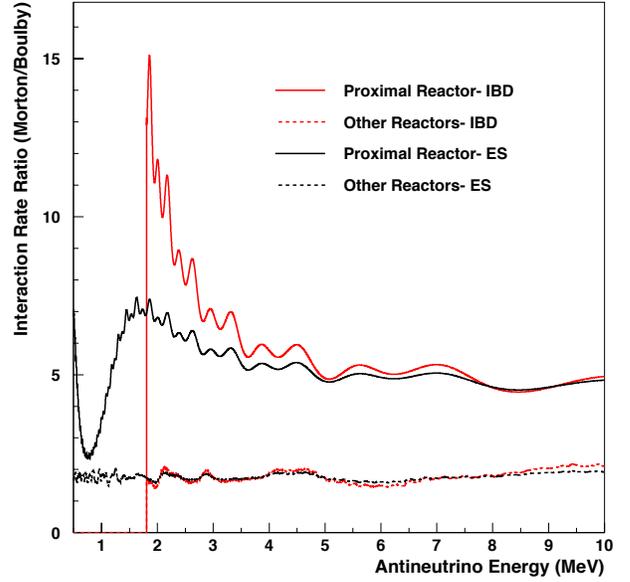}
\caption{Ratios of the quasi-elastic (IBD) and elastic (ES) interaction energy spectra at the two sites from both the proximal reactor and all other reactors.}
\label{fig:site_ratio}
\end{figure}

The number $N(L,E_\nu)$ of reactor antineutrino interactions as a function of standoff distance $L$ and antineutrino energy $E_\nu$ at a given site follows from
\begin{equation}
N(L,E_\nu)=\frac{n\tau}{4\pi L^2}\int\sigma(E_\nu)\frac{dR}{dE_\nu}P(L,E_\nu)dE_\nu,
\end{equation}
where $n$ is the number of targets (for \eqref{ibd} free protons or hydrogen nuclei and for \eqref{es} atomic electrons) and $\tau$ is the exposure time. For the present study a convenient unit of exposure corresponds to a detector with a target mass of $1$ kT of water, where $n_p=6.69\times10^{31}$ for \eqref{pxsec} and $n_e=3.35\times10^{32}$ for \eqref{exsec} and \eqref{muxsec}, that is operated for $1$ year, where $\tau=3.16\times10^7$ s. Figure~\ref{fig:ibd_rate} compares the calculated spectra for reaction \eqref{ibd} at the two reactor-site combinations. Figure~\ref{fig:es_rate} compares the calculated spectra for reaction \eqref{es} from the proximal reactor at the two sites. Figure~\ref{fig:es_rate_other} compares the calculated spectra for reaction \eqref{es} from all other reactors at the two sites. While the signal rate from the proximal reactor is about a factor of five higher at Morton than at Boulby, the background rate from all other reactors is only about a factor of two higher at Morton than at Boulby. Figure~\ref{fig:site_ratio} shows the spectra of these rate ratios, indicating that the background fraction is higher at Boulby than at Morton.

The calculations here estimate 6157 (1330) IBD (ES) interactions per kT-y for Perry-Morton and 1048 (242) IBD (ES) interactions per kT-y for Hartlepool-Boulby. These estimates apply to the proximal reactor cores operating continuously at full power and perfect detection efficiency. The Perry-Morton signal rates are more than five times greater than the Hartlepool-Boulby signal rates due to shorter reactor-site separation distance, larger reactor thermal power, and higher neutrino oscillation survival probability. 

\begin{figure}
\centering
\includegraphics[trim = 5mm 40mm 15mm 40mm, clip, scale=0.45]{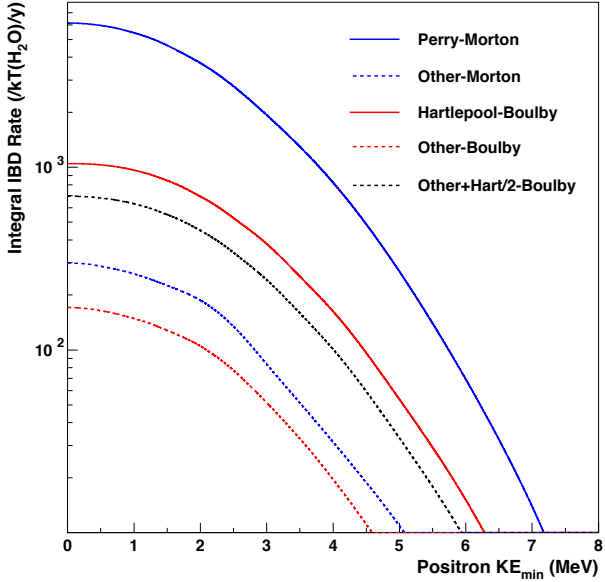}
\caption{Integral rate of quasi-elastic (IBD) interactions as a function of positron minimum kinetic energy from both the proximal reactor and all other reactors at the Morton and Boulby sites. The dashed black curve sums the integral rate of quasi-elastic interactions from one of the cores at Hartlepool (Hart/2) and all other reactors at the Boulby site.}
\label{fig:ibd_rate_int}
\end{figure}

\begin{figure}
\centering
\includegraphics[trim = 5mm 40mm 15mm 40mm, clip, scale=0.45]{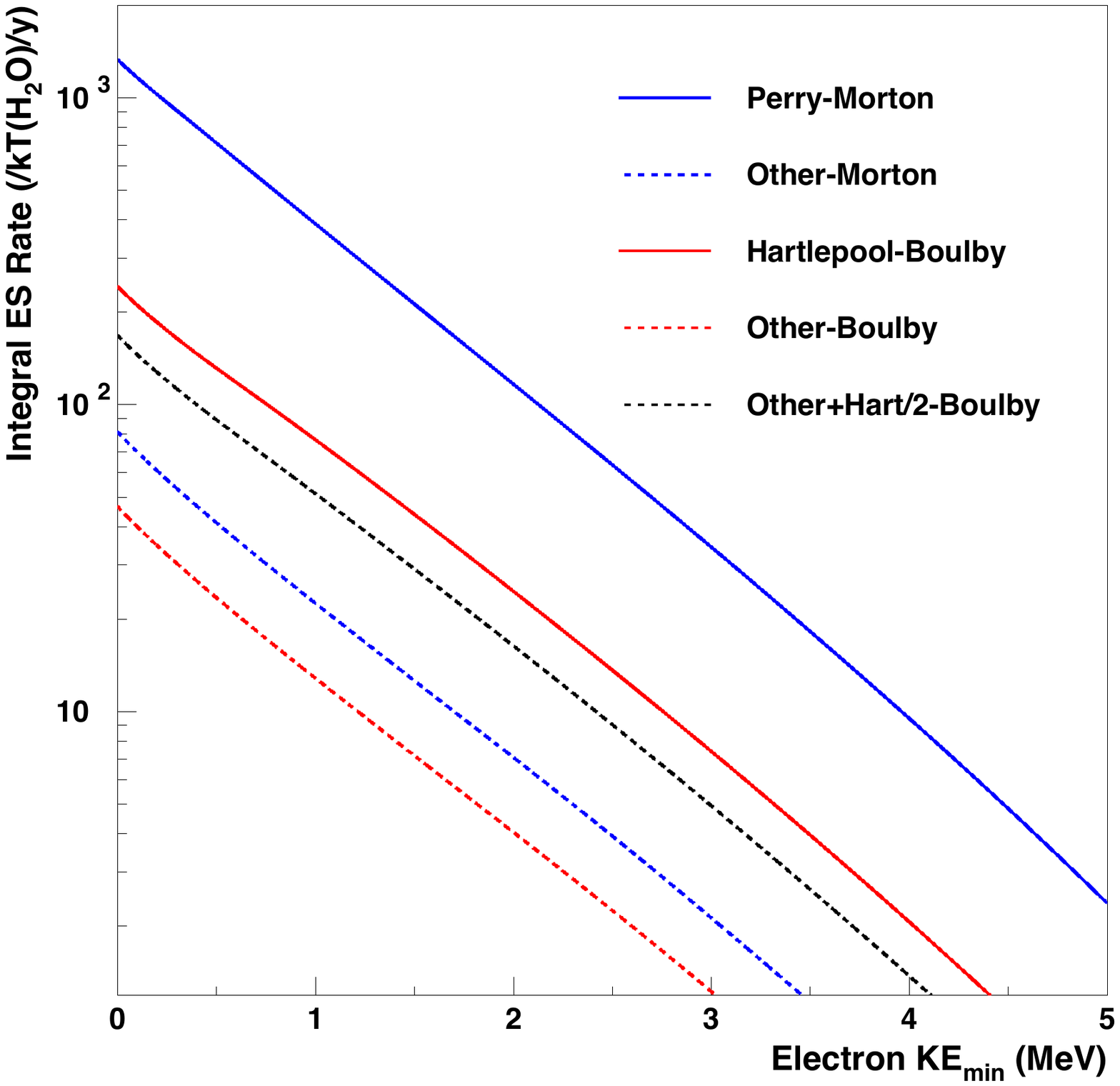}
\caption{Integral rate of elastic (ES) interactions as a function of electron minimum kinetic energy from both the proximal reactor and all other reactors at the Morton and Boulby sites.The dashed black curve sums the integral rate of quasi-elastic interactions from one of the cores at Hartlepool (Hart/2) and all other reactors at the Boulby site.}
\label{fig:es_rate_int}
\end{figure}

The scattered positron in \eqref{ibd} and electron in \eqref{es} produce the detectable signal. Figure~\ref{fig:ibd_rate_int} shows the integral number of quasi-elastic scattering interactions per kT-y exposure of water above a given positron kinetic energy for the two reactor-site combinations. Figure~\ref{fig:es_rate_int} shows the integral number of elastic scattering interactions per kT-y exposure of water above a given electron kinetic energy for the two reactor-site combinations. Both Fig. \ref{fig:ibd_rate_int} and Fig. \ref{fig:es_rate_int} show separately the signal due to the proximal reactor (Perry or Hartlepool) operating at full power and the background from all other reactors operating at their annual average load factor during the year 2014 \cite{infnsite}. Table~\ref{tab:ibdsum} presents the integral rates of quasi-elastic interactions for selected positron kinetic energy thresholds. Table~\ref{tab:essum} presents the integral rates of elastic interactions for selected electron kinetic energy thresholds. Because the Hartlepool reactor has two cores, the figures display the spectrum representing one-half the signal from Hartlepool added to the spectrum from all other reactors. Presumably this is the background at the Boulby site for the reactor on/off study.

\begin{table}
\caption{Quasi-elastic scattering interaction rates per (kT-y) exposure of water for various positron kinetic energy thresholds estimated for both the proximal reactor and all other reactors at the Morton and Boulby sites.}

\begin{tabular}{l r r r r r r}
\hline\noalign{\smallskip}
Minimum $T_e$ (MeV)  & 0.0 & 0.7 & 1.5 & 2.5 & 3.5 & 4.5 \\
\hline\noalign{\smallskip}
Perry-Morton         & 6157 & 5800 & 4641 & 2776 & 1296 & 488 \\
Hartlepool-Boulby & 1048 & 1010  & 846 & 528 & 253 & 96.5 \\
\hline
Other-Morton         & 299 & 278 & 225 & 135 & 51.4 &  18.8 \\
Other-Boulby        & 171 & 159 & 129 & 77.4 & 32.7 &  11.1 \\
\hline\noalign{\smallskip}
\end{tabular}
\label{tab:ibdsum}
\end{table}

\begin{figure}
\centering
\includegraphics[trim = 5mm 40mm 15mm 40mm, clip, scale=0.45]{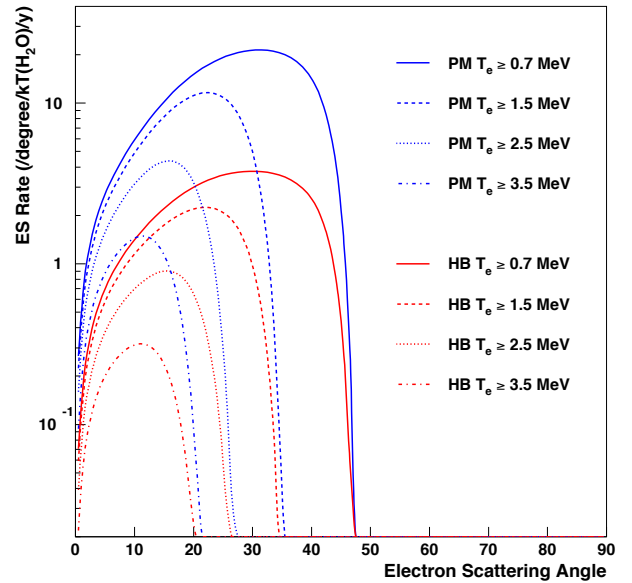}
\caption{Spectra of electron scattering angles from ES interactions due to the proximal reactor at the Morton (PM) and Boulby (HB) sites.}
\label{fig:ang_spec}
\end{figure}

An attribute of the elastic scattering process is the potential for estimating the antineutrino direction from the measured electron track. The emission angle of Cherenkov radiation is $\cos\theta=1/(\beta n)$, where $\beta=v/c$ and $n$ is the index of refraction of the medium. In water, emission is at an angle of $\simeq42^{\circ}$ for $\beta=1$ and emission stops when $\beta\lesssim0.75$. An electron with kinetic energy $\lesssim0.26$ MeV does not produce Cherenkov radiation in water. The distribution of scattering angles clusters in the forward direction as the kinetic energy of the electron increases. Due to the steeply falling reactor antineutrino energy spectrum the number of interactions diminishes with increasing electron kinetic energy. Figure~\ref{fig:ang_spec} shows the spectra of electron scattering angles for selected kinetic energies above Cherenkov threshold. 

\begin{table}
\caption{Elastic scattering interaction rates per (kT-y) exposure of water for various electron kinetic energy thresholds estimated for both the proximal reactor and all other reactors at the Morton and Boulby sites.}

\begin{tabular}{l r r r r r r}
\hline\noalign{\smallskip}
Minimum $T_e$ (MeV)  & 0.0 & 0.7 & 1.5 & 2.5 & 3.5 & 4.5 \\
\hline\noalign{\smallskip}
Perry-Morton         & 1330 & 558 & 213 & 63.6 & 18.2 & 4.8 \\
Hartlepool-Boulby & 242 & 106  & 43.9 & 13.6 & 4.0 & 1.0 \\
\hline
Other-Morton         & 81.6 & 32.2 & 12.6 & 3.93 & 1.13 &  1.05 \\
Other-Boulby        & 46.5 & 18.4 & 7.18 & 2.25 & 0.65 &  0.18 \\
\hline\noalign{\smallskip}
\end{tabular}
\label{tab:essum}
\end{table} 

\section{Discussion}
The reactor antineutrino interaction rates and spectra calculated herein contribute to assessments of the nuclear monitoring capabilities of the Morton and Boulby detector sites. Full assessments require a more complete accounting of background sources, including geological antineutrinos, solar neutrinos, cosmogenic nuclides, neutrons, and detector noise, including radon contamination. Combined samples of signal and background then need to pass through the detector simulation and data analysis routines. The immediate goal of this report is to initiate the process of evaluating the potential of the Morton and Boulby sites to fulfill the WATCHMAN nuclear monitoring agenda \cite{watchman}. 

\section{Conclusions}
This report presents calculations of the reactor antineutrino interactions, from both quasi-elastic neutrino-proton scattering and elastic neutrino-electron scattering, in a water-based detector operated $\gtrsim10$ km from a commercial power reactor. It separately calculates signal from the proximal reactor and background from all other registered reactors. The main results are the interaction rates and kinetic energy distributions of the charged leptons scattered from the quasi-elastic and elastic processes. Comparing signal and background distributions evaluates reactor monitoring capability. Scaling the results to detectors of different sizes, target media, and standoff distances is straightforward. Calculations are for two specific examples of a commercial reactor ($P_{th}\sim3$ GW) operating nearby ($L\sim20$ km) an underground facility capable of hosting a detector ($\sim1$ kT H$_2$O) project. These reactor-site combinations are Perry-Morton on the southern shore of Lake Erie in the United States and Hartlepool-Boulby on the western shore of the North Sea in England. The signal from the proximal reactor is about five times greater at the Morton site than at the Boulby site due to shorter reactor-site separation distance, larger reactor thermal power, and greater neutrino oscillation survival probability. In terms of absolute interaction rate, background from all other reactors is larger at Morton than at Boulby. However, the fraction of the total rate is smaller at Morton than at Boulby. Moreover, the Hartlepool power plant has two cores whereas the Perry plant has a single core. These conditions make monitoring the operation cycle of a nuclear reactor more challenging at the Boulby site than at the Morton site. The Boulby site, therefore, offers an opportunity for demonstrating remote reactor monitoring under more stringent conditions than does the Morton site.

\section*{Acknowledgments}
This work was supported in part by Lawrence Livermore National Laboratory and the Trustees' Scholarly Endeavors Program at Hawaii Pacific University.

\newpage
\bibliographystyle{apsrev4-1}
\nocite{apsrev41control}
\bibliography{Nureac_bib,revtex-custom}

\end{document}